  \documentstyle[epsfig]{article}

\begin{document}
%
%
%
%

\begin{center}
{\Large \bf Studies of a weak polyampholyte}\\ \vskip 0.2cm
{\Large \bf at the air-buffer interface:}\vskip 0.3cm
 {\Large \bf The effect of varying pH and ionic strength}
 \vskip 1cm
 {\large { Pietro Cicuta, Ian Hopkinson\footnote{E-mail address:
Ian.Hopkinson@phy.cam.ac.uk}} } \vskip 0.2cm

Cavendish Laboratory\\ University of Cambridge\\ Madingley Road,
Cambridge CB3 0HE, U.K.
\end{center}

\vskip 0.1 cm \vspace{1 cm}

%
%
%
\begin{abstract}
We have carried out experiments to probe the static and dynamic
interfacial properties of $\beta$--casein monolayers spread at the
air-buffer interface, and analysed these results in the context of
models of weak polyampholytes. Measurements have been made
systematically over a wide range of ionic strength and pH. In the
semi-dilute regime of surface concentration a scaling exponent,
which can be linked to the degree of chain swelling, is found.
This shows that at pH close to the isoelectric point, the protein
is compact. At pH away from the isoelectric pH the protein is
extended. The transition between compact and extended states is
continuous. As a function of increasing ionic strength, we observe
swelling of the protein at the isoelectric pH but contraction of
the protein at pH values away from it. These behaviours are
typical of a those predicted theoretically for a weak
polyampholyte. Dilational moduli measurements, made as a function
of surface concentration exhibit maxima that are linked to the
collapse of hydrophilic regions of the protein into the subphase.
Based on this data we present a configuration map of the protein
configuration in the monolayer. These findings are supported by
strain (surface pressure) relaxation measurements and surface
quasi-elastic light scattering (SQELS) measurements which suggest
the existence of loops and tails in the subphase at higher surface
concentrations.
\end{abstract}
%
%
%
%
 \vskip 2cm

\newpage
\section{Introduction}
A polyampholyte is a polymer which contains both positively and
negatively charged monomers \cite{Kud99}. A weak polyampholyte is
one where the overall charge can be adjusted by varying external
conditions, usually the pH. This can lead to overall neutral
polyampholytes, whose net charge is zero, or polyampholytes which
have a net positive or negative charge. The behaviour of such
molecules is of interest because amongst their number are found
the proteins. An understanding of the way in which polyampholytes
behave, in particular the transition from an open coil to a
compact globule, may give some insight to the understanding of
protein folding.

 Polyampholytes are also of interest industrially. Polymers with a
 net charge,
polyelectrolytes, are often used as stabilisers, thickeners and in
oil recovery. However in situations of high salinity, commonly
found in oil recovery, their thickening effect declines
dramatically. Polyampholytes hold out the prospect of overcoming
this limitation.\\

There have been a number of theoretical approaches to analysing
polyampholytes, these include scaling arguments \cite{HJ91}, Monte
Carlo simulations \cite{LT00}, \cite{KK95}, \cite{LO98},
\cite{Yam00}, molecular dynamics \cite{TT00}, Flory theory models
\cite{DR95} and analogies with charged droplets \cite{KK95}.
Previous experimental studies on polyampholytes have looked at the
swelling of gels \cite{NMC99},\cite{Eng96}, \cite{CC93} or the
changes in viscosity of polyampholyte solutions \cite{CC93}.

The study of polymers at air-liquid and liquid-liquid interfaces
is of technological and academic interest \cite{JR99},
\cite{Ada97}, \cite{Mob98}. This is a result of the importance of
polymers at interfaces as stabilisers in multiphase systems and as
modifiers of interfacial properties. More recently attention has
turned towards charged polymers, which introduce added complexity
through electrostatic interactions. This is in part through a
desire to understand the properties of natural polymers, such as
proteins and polysaccharides, which are usually charged.

Close to overall neutrality, polyampholytes collapse to a compact
configuration. This is because the chain can rearrange in order to
facilitate attractive opposite-charge interactions which leads to
a compact state. Above a net charge eN$^{1/2}$, where N is the
number of monomers and e is the charge density, the chains swell.
This is because, as the net charge on the polyampholyte increases,
it becomes increasingly difficult to rearrange to a configuration
that allows opposite charge attractions without incurring an
energy penalty through the now more numerous like charge
repulsions. In three dimensional synthetic polyampholyte gels the
swelling transition is abrupt. As a function of ionic strength,
gels with a net charge are found to be compact at very low ionic
strength, swelling to a maximum then collapsing. At very high
ionic strength they swell again, this arises from the osmotic
pressure of a large number of counterions condensed onto the
polyampholyte. Analogies with charged liquid droplets and Monte
Carlo simulations suggest that, at overall charges beyond the
eN$^{1/2}$ threshold, polyampholytes will break up into a ``string
of pearls'', where the number of ``pearls'' is ca. Q/Q$_c$, where
Q$_c$ is the threshold charge described above and Q is the net
charge. At high ionic strengths polyampholytes behave like
polyelectrolytes through the screening of charge by counterion
condensation.

Measurements of
surface~pressure~($\Pi$)--surface~concentration~($\Gamma$) curves
of interfacial layers are the two dimensional equivalent of
pressure--volume curves and in a similar manner can provide
information on the inter- and intra-molecular interactions
controlling the properties of the layer. At low surface
concentrations molecules move independently in the so-called
dilute regime. There is a  surface concentration $\Gamma=\Gamma^*$
beyond which polymer molecules  start to overlap, this marks the
beginning of the ``semi--dilute'' regime. The behaviour of the
$\Pi$--$\Gamma$ isotherm in the semi--dilute regime of a polymer
layer can be used to measure the Flory--Huggins exponent
\cite{DJ76} which is related to the relative 'compactness' of a
polymer molecule. Measurements of interfacial strain relaxation
can be made which, in common with comparable bulk measurements,
can give an insight into molecular relaxations.

In this work in addition to using 'zero frequency' interfacial
techniques, we have also applied surface quasi--elastic light
scattering (SQELS) to probe the interfacial properties at high
frequency. Langevin \cite{Lan92} gives an excellent review of the
technique and its applications. Surface quasi--elastic light
scattering measures the power spectrum of the thermally driven
fluctuations of a fluid interface, and from this information the
viscoelastic properties of that interface can be determined. The
measured thermally driven fluctuations typically have wavelengths
of the order of 100$\mu$m, amplitudes of the order 2$\AA$ and
frequencies of the order 10kHz. Here we use SQELS for two
particular reasons. Firstly SQELS probes the interfacial
properties at high frequencies compared to any other conventional
method, thus it is relevant to rapid events found in processes
such as emulsification. Secondly we can obtain a dilational
viscosity, $\epsilon '$, which is not otherwise measurable in this
frequency regime. This dilational viscosity is helpful in
validating models for the behaviour of the surface dilational
modulus with frequency.

In this work $\beta$--casein was used as a model, weak
polyampholyte. $\beta$--casein is a milk protein with a random
coil structure and a molecular weight of ca.~24kDa. It is used
extensively in the food industry as an `emulsifier'. Its
emulsifying properties are believed to arise from the blocky
distribution of hydrophilic and hydrophobic residues along its
length. We will discuss the amino acid sequence in more detail
below, but essentially the first 50 residues at the N-terminal end
of the protein are rather more hydrophilic than the rest of the
molecule.

Neutron reflectometry  \cite{Dic93} and ellipsometry experiments
\cite{GP79} , \cite{HKC91} have been performed on $\beta$--casein
at high surface coverages. They have shown that the protein
adsorbs to the interface with a thin dense layer right at the
surface and a thicker, less dense layer beneath it. It has been
proposed that this sublayer is composed of the hydrophilic
N-terminal end of the molecule. This picture is supported by
proteolytic cleavage experiments \cite{Mel98} which show the loss
of this terminal region when the adsorbed protein is exposed to a
cleavage enzyme. Monte Carlo simulations \cite{APR00} and
self-consistent field simulations \cite{Lee96} have been carried
out using monomer sequences which replicate $\beta$--casein on a
coarse grain level. These simulations are in agreement with the
experimental observations, showing a dense surface layer and a
less dense sublayer corresponding to hydrophilic regions of the
protein.

The experiments presented here complement these studies because
they focus on a lower surface concentration, where neutron
experiments have not been done. Furthermore, we have looked
systematically at the behaviour of $\beta$--casein on buffers of a
wide range of pH and ionic strength.

Douillard \cite{Dou94} and Aguié-Béghin \cite{Agu99} have used
scaling arguments to model the surface~pressure--concentration
isotherm of protein layers and interfacialy adsorbed multiblock
copolymers. These models provide a general framework which is not
inconsistent with the behaviour of $\beta$--casein, however they
do not directly account for changes in behaviour with pH and ionic
strength. Fainerman and Miller \cite{FM99} have made predictions
of the shape of the isotherm at high $\Gamma$ by considering the
thermodynamics of aggregates and their equilibrium in the surface
layer.

We have calculatedthe likely charge distribution on the protein
from the published primary sequence \cite{BA00} and the known
dissociation constants for amino acids. This calculation shows
overall neutrality at around pH=5, in agreement with the published
isoelectric point \cite{Swa82}. An overall positive net charge
arises below this pH, and an overall negative charge above it. We
estimate  the net charge to be around -10 at pH=9. The total
number of charges on the molecule is between 30 and 50 throughout
the investigated range of pH. We observe the well known
hydrophilic tail at the N-terminal, and in addition tentatively
identify two further hydrophilic regions. One of these is more
substantial and lies in the middle of the chain, the other lies
near the C--terminal end.

There are a number of key parameters important in the study of
solutions containing charged molecules. The Bjerrum length, l$_B$,
is the distance from a charge at which electrostatic and thermal
energies are comparable:
\begin{eqnarray}
 l_B\,=\,\frac{e^2}{4\,\pi\,\epsilon_r\,\epsilon_0\,k_B\,T}
 \label{eq1}
\end{eqnarray}
where e is the electron charge, $\epsilon_r$ is the relative
permittivity of water, $\epsilon_0$ is the permittivity of vacuum,
$k_B$ is the Boltzmann constant and T is the temperature. In water
at room temperature the Bjerrum length is 7.14$\AA$. (Note that
definitions in the literature appear to vary, but this is normally
because factors of 4$\pi$ are required in converting between CGS
and SI units). The Debye length, $\kappa^{-1}$, gives a measure of
the distance at which electrostatic interactions are screened out
by the presence of ions in solution:
\begin{eqnarray}
 \kappa^{-1}\,=\,(8\pi\,n\,l_B)^{-\frac{1}{2}}
 \label{eq2}
\end{eqnarray}
where, n is the number density of charges. In the NaCl solutions
used here, r$_D$=3.035$\AA$/I$^{1/2}$, where I is the molar ionic
strength. In this work the Debye length varies from 96$\AA$ at
I=0.001M to 3$\AA$ at I=1.1M. A final quantity of interest is the
'effective' temperature, t. This is given by \cite{DR95}:
\begin{eqnarray}
 t\,=\,\frac{b\,N}{l_B\,(N^+\,+\,N^-)}
 \label{eq3}
\end{eqnarray}
where N is the number of monomers per chain, N$^+$ and N$^-$ are
the numbers of positive and negative charges on the chain. b is
the statistical segment length of the protein. For $\beta$--casein
in guanidinium chloride (Gdn-HCl) solution the radius of gyration,
Rg, measured by neutron scattering is 69$\AA$ \cite{Cal95},
therefore b=11.7$\AA$ (since R$_g$=(N/6)$^{1/2}$b). Clearly we
would expect the radius of gyration to vary with solution
conditions, we simply use these data to show that l$_B$ and b are
of the same magnitude. In these experiments (\,N$^+$+ N$^-$\,)
lies in the range 30-50, therefore the system is always in the
'high effective temperature' limit. In order to reach the low
temperature limit it is necessary to use a very highly charged
molecule.\\

\section{Experimental Methods}
\subsection{Materials} $\beta$--casein (Sigma, C-6905, 90\% pure) was used as
supplied. 1mg/ml solutions in deionised water were prepared from
the dried, powdered protein, stored in a refrigerator and used
within 5 days. Buffer solutions were made up using deionised
(Elgastat UHQ, Elga, U.K.) water. Buffers with a range of pH and
ionic strength were prepared. For pH in the range 5.8 to 8.5
phosphate buffer was used, below pH=5.8 citrate buffer was used
and above pH=8.5 carbonate buffer \cite{Daw59}. To control ionic
strength NaCl was added, quoted ionic strengths also include the
contribution of the buffer salts. Buffer pH were measured using an
electronic meter (ATI Orion, USA) before use.

\subsection{Langmuir trough methods}
 Surface pressure vs. area isotherms were measured using a Langmuir
 trough with a filter paper Wilhelmy
plate sensor (Nima Technology, U.K.) mounted on an active
anti-vibration table (Halycion, Germany), both of which were
enclosed in a draft proof enclosure. These procedures, which
reduce extraneous vibration, are necessary in order to carry out
the SQELS measurements described below. The PTFE trough, area
530cm$^2$ was filled with approximately 500ml of the appropriate
buffer. Surface pressure - area isotherms of the 'bare' buffer
were measured before each experiment. The surface was aspirated
and the isotherm re-measured until it showed no increase in
surface pressure on full compression. This was to ensure there was
no surface contamination prior to the addition of the
$\beta$--casein. $\beta$--casein was spread on the surface of the
buffer by careful dropwise addition of the appropriate volume of
the 1mg/ml protein solution. The initial volume of added protein
solution was chosen such that the surface layer was initially in
the dilute regime. Typically the volume dispensed was
ca.~25$\mu$l. Lower concentrations of spreading solution were
found to lead to loss into the subphase on spreading. Repeated
spreading was used to explore a greater concentration range than
possible with a single compression. Since hysteresis is observed
upon expansion following compression to high pressures
($\Pi>10\times 10^{-3}N/m$), no data is presented from
compressions of monolayers having such a history. These  methods
has been validated for $\beta$--casein in \cite{NSP99}. The
temperature of the trough was held at 22$^o$C by running water
from a temperature controlled water bath (Haake, Germany) through
channels in the base of the Langmuir trough. A header tank was
used to avoid transmitting vibrations to the liquid surface.\\
Three types of measurement were made:\\ (1) Surface
pressure($\Pi$)-surface concentration($\Gamma$) isotherms, the
surface pressure was measured as the trough barrier was moved at a
constant compression rate of 40cm$^2$/min.\\ (2) Step compression
measurements, the surface area, A, was changed by a constant
amount ($\Delta$A/A = 0.55) at a barrier speed of 1800cm$^2$/min.
The subsequent evolution of the surface pressure was measured for
15 minutes. The trough area was then further reduced and the next
relaxation
 measured.\\ (3) SQELS measurements were made during the
relaxation phase of the step compression experiments. The
magnitude of the pressure relaxation is small when compared to the
accuracy of the SQELS measurement. Experiments were all carried
out within 3-4 hours of the protein first being spread on the
buffer. We do not observe aging effects within this time period.
We are aware that over a period of many hours and even days,
particularly for adsorbed monolayers, such aging has been
observed.

\subsection{Surface quasi-elastic light scattering}
\begin{figure}[ht]
          \begin{center}
           \epsfig{file=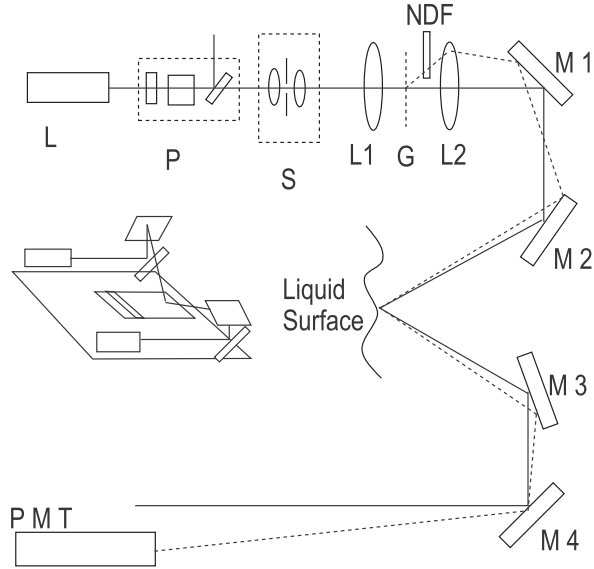,width=8.5cm}
          \end{center}
 \caption{Schematics of the SQELS apparatus, with inset
showing an isometric view. Symbols are explained in the
text.}\label{F3}
\end{figure}

 We use a surface quasi-elastic light scattering (SQELS) apparatus, built in house,
based on a design proposed by Earnshaw \cite{EM87} and H{\aa}rd
\cite{HN81}, this is illustrated schematically in Fig.~\ref{F3}.
%
%
%
%
 The goal of such an apparatus is to measure the power
spectrum of light scattered inelastically from the capillary waves
at the fluid interface as a function of scattering vector, q,
measured relative to the specular reflection.  Photon correlation
spectroscopy (PCS) is a convenient means by which to measure the
small shifts in frequency that this entails. The photon
correlation is done in heterodyne mode and so it is necessary to
provide a coherent source of light of the original frequency at
the appropriate q value. This light is provided using a weak
diffraction grating. In order for the heterodyne signal to
dominate the correlation function, the ratio of the intensity of
the inelastically scattered light to the 'reference' light must be
adjusted to a value of the order of 10$^{-3}$.\\ Turning to
Fig.~\ref{F3}: Light, with wavelength of 532nm, is provided by a
150mW single mode diode pumped solid state laser (Laser Quantum,
U.K.). Polarisation and intensity are controlled using the
combination of the half wave plate ($\lambda$/2) and prism
polariser (P). The beam size, profile and collimation are
controlled using the spatial filter, S. The grating (G) provides a
fan of diffracted 'reference' beams. The lenses L1 (f=150mm) and
L2 (f=350mm) perform two tasks; they converge the reference beams
and the main beam to a single spot at the fluid interface and they
focus the reference beams and the main beam in the front plane of
the photomultiplier, situated ca.~2m after the surface. The
relative intensity of the reference beams is adjusted by inserting
a neutral density filter (NDF) so that it intercepts the
diffracted spots but not the main beam. The mirrors M1-M4 direct
light from the laser onto the surface and from there into the
detector. The light is detected using a photomultiplier (PMT) and
processed using a PC-card based photon correlator (BI9000,
Brookhaven Instruments, USA), the pulse discriminator used in the
PMT is modified to allow the use of the 'multi-photon' mode
originally described by Earnshaw \cite{WE88}. At the detector the
laser light appears as a bright central spot with a series of
focussed reference spots at 2-3mm intervals away from the central
spot. Each of these spots is composed of the reference beam
originating from the diffraction grating and inelastically
scattered light from the main beam. The reference beams are
sufficiently weak that inelastic scatter from them can be ignored.
Each spot corresponds to light being scattered to a different q
value, the mirror M4 is adjusted in order that the appropriate
reference beam falls on the detector. The value for the scattering
vector, q, and the instrument resolution at a particularly
reference spot was calibrated by fitting the measured correlation
function for a pure liquid (water) at that point. Quoted values of
the interfacial properties are the mean of values derived from
fitting 5 correlation functions acquired consecutively under the
same conditions, with each correlation function accumulated over
two minutes. To reduce stray light, a mirror on the trough bottom
deflects away from the forward direction any light that had not
been specularly reflected by the buffer interface.

SQELS data were acquired from protein decorated air-buffer
interfaces maintained in a Langmuir trough as described above.
Data for a particular surface concentration were acquired with the
barrier of the trough stationary, although we have found it
possible to acquire data from a protein layer undergoing very slow
compression.\\

\subsection{Data Analysis Methods}
\subsubsection{Dilational Moduli}
  The surface pressure, $\Pi$,
of a protein layer is the difference between the surface energy
per unit area of the bare buffer, $\gamma_0$, and the surface
energy measured with the layer in place, $\gamma$. Features in the
$\Pi$--$\Gamma$ isotherms are seen more clearly if the dilational
modulus, $\epsilon_{st}$, is calculated from the isotherm, and
this is then plotted as a function of $\Gamma$. For an insoluble
layer:
\begin{eqnarray}
 \epsilon_{st}\,=\,\Gamma\,\frac{d\Pi}{d\Gamma}
 \label{eq4}
\end{eqnarray}
The dilational modulus is the in--plane 'dilational' elasticity of
the surface layer. For uniaxial stress, as found in the
measurements made here, the dilational modulus is the sum of the
compressional and shear moduli \cite{Goo81}.\\
\subsubsection{Scaling Exponents} The $\Pi$--$\Gamma$ isotherms
can be described in the semi-dilute regime using a scaling law,
with exponent y:
\begin{eqnarray}
 \Pi\,\propto\,\Gamma^y
 \label{eq5}
\end{eqnarray}
This was introduced by Daoud,  Jannik and de Gennes \cite{DJ76},
\cite{DG77}, first verified for a polymer monolayer by Vilanove
\cite{Vil?} and has since been applied to a wide range of polymer
monolayers \cite{JR99}. Douillard \cite{Dou94} and
Agui\'{e}-B\'{e}ghin \cite{Agu99} have further developed these
ideas to apply to multiblock copolymers. The Flory scaling
exponent, $\nu$, relating the chain radius of gyration, R$_g$, to
the number of monomers, N,
\begin{eqnarray}
R_g\,\propto\,N^{\nu}
 \label{eq6}
\end{eqnarray}
is connected to y by y = 2$\nu$/(2$\nu$-1) in 2D, and y =
3$\nu$/(3$\nu$-1) in 3D. The value of the exponents are different
in 2D and 3D chains, and expected values are summarised in Table 1
for different solvent conditions.\\
%
%
%
%
Table 1:\\ Values of the two dimensional ($\nu$2) and three
dimensional ($\nu$3) theoretically calculated Flory exponents, and
values of the corresponding (y2 and y3)  exponents y
($\Pi\propto\Gamma^y$ in the semi--dilute regime), for different
solvent conditions.\\

\begin{center}
\begin{tabular}{|c||c|c|c|c|}  \hline
    Conditions  &   \,$\nu$2\,   &        \,$\nu$3\, & \,y2\, & \,y3\, \\
    \hline \hline
     Extended Chain & \,1 \,   &        \,1 \, &  \,2 \, &  \,3/2 \, \\ \hline
     Good Solvent &  \,3/4 \, &        \,3/5 \, &  \,3 \, &  \,9/4 \, \\ \hline
          $\theta$ solvent &  \,4/7 \, &  \,1/2 \, &  \,8 \, &  \,3 \, \\ \hline
        Poor Solvent  &  \,1/2 \,   &   \,1/3 \, &  \,$\infty$ \, &  \,$\infty$ \, \\ \hline
\end{tabular}
\end{center}
 In 2D y is
small (\,y=3\,) for good solvents and increases to y=8 as solvent
quality moves towards the $\theta$ conditions. Here we use the
scaling exponent as a measure of the overall compactness of the
protein rather than trying to link observed behaviour to very
specific changes. We can measure the scaling exponent from the
slope of the log-log plot of $\Pi$--$\Gamma$. However the slope of
the $\epsilon$--$\Pi$ plot \cite{Agu99} also recovers the exponent
y and this method is preferred because it is not sensitive to
errors in the amount of spread solution. We find in all conditions
that the $\epsilon$--$\Pi$ plot is linear in, at least, the
surface pressure range $0-2\,\times10^{-3}$N/m.
  \subsubsection{Surface
quasi-elastic light scattering} 
\begin{figure}[h!]
         \begin{center}
         \epsfig{file=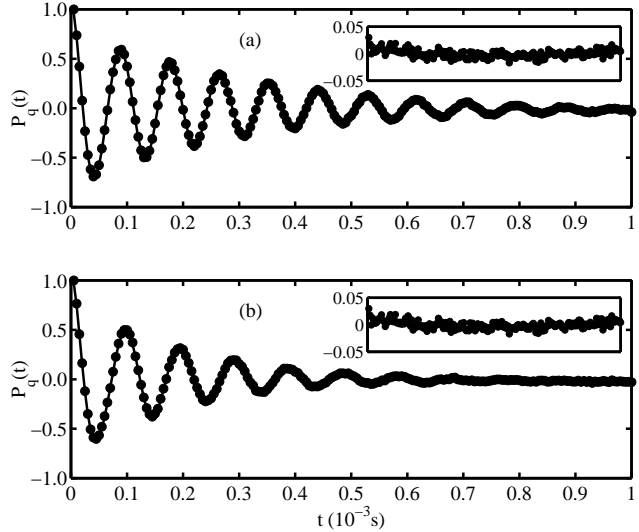,width=8.5cm}
       \end{center}
\label{F4} \caption{Typical correlation functions obtained by
surface quasi--elastic light scattering  at subphase pH=5.24 and
scattering vector q=425cm$^{-1}$ under different conditions (a):
`bare' buffer; (b): $\beta$--casein monolayer at concentration
$\Gamma=1 \times 10^{-3}g/m^2$ and pressure $\Pi=6.3\times
10^{-3}N/m$. The solid lines are fits with the model described in
the text, and insets show the residuals of the fits.}
\end{figure}

SQELS data are normally analysed
in terms of a model treating the interfacial layer as a thin flat
elastic sheet at the interface \cite{Ear90} . More recently Buzza
et al.\cite{Buz98}  have proposed a model that explicitly
incorporates features of a polymer brush into the model of the
interfacial layer, introducing bending and coupling moduli.
Numerical analysis showed that the Buzza model would reduce to a
thin viscoelastic sheet for the interfacial layer thickness
measured for $\beta$--casein. We base our presentation of the key
results for the analysis of SQELS data on the paper by Earnshaw et
al  \cite{Ear90} and refer to the paper by Buzza et al
\cite{Buz98} for some clarification and issues related
specifically to polymer monolayers. The dispersion relation
D($\omega$) for waves at an air-liquid interface, bearing a thin
viscoelastic layer, is given by
\begin{eqnarray}
D(\omega)\,=\,\left[\,\epsilon q^2\,+\,i\omega \eta \left(
q\,+\,m\right)\, \right]\,\left[\,\gamma q^2\,+\,i\omega\eta
\left( q\,+\,m \right)\,-\,\frac{\rho \omega^2}{q}\,
\right]\,-\,\left[\,i\omega \eta \left(m\,-\,q \right)\,
\right]^2,
 \label{eq7}
\end{eqnarray}
where m is
\begin{eqnarray}
m\,=\,\sqrt{\,q^2\,+\,i\frac{\omega \rho}{\eta}}, \,\,\,Re(m)>0,
 \label{eq8}
\end{eqnarray}
$\eta$ is the subphase viscosity, $\rho$ is the subphase density,
$\gamma$ is the surface tension (or transverse modulus) and
$\epsilon$ is the dilational modulus.\\ Solving this equation for
D($\omega$)\,=\,0 gives us an expression for the wave frequency,
$\omega$, as a function of the scattering vector q. The solutions
describe both dilational and transverse waves. In a light
scattering experiment it is only the transverse waves that scatter
light and their power spectrum P$_q$($\omega$) is given by:
\begin{eqnarray}
P_q(\omega)\,=\,\frac{k_B T}{\pi \omega}Im\left[\,\frac{i \omega
\eta (m\,+\,q)\,+\,\epsilon q^2 }{D(\omega)}\, \right].
 \label{eq9}
\end{eqnarray}
The behaviour of the dilational waves can be inferred because of
their coupling to the transverse waves. A fluid--fluid interface
can be modelled using a trivial modification of~(\ref{eq7}). In
the experiments carried out here a photon correlation spectrum is
acquired. After accounting for instrumental factors this is simply
the time Fourier transform, P$_q$(t) of the power spectrum,
P$_q$($\omega$). The dilational modulus can be expanded to take
into account viscous effects:
\begin{eqnarray}
\epsilon\,=\,\epsilon_0\,+\,i \omega \epsilon ',
 \label{eq10}
\end{eqnarray}
where $\epsilon_0$ is the dilational modulus and $\epsilon$' is
the dilational viscosity. Buzza et al have shown that such an
expansion is not appropriate for the surface tension, and
$\gamma$' should be set to zero. In this work data is analysed by
directly fitting the measured correlation function with a
correlation function, P$_q$(t), curve calculated from the
interfacial properties \cite{Ear90}. The interfacial parameters
$\epsilon$, $\epsilon$' and $\gamma$ are all fitted
simultaneously. An example of raw  P$_q$(t) correlation data
fitted with a modelled function as described above is shown in
Fig.~\ref{F4}.
%
%
%
%

An alternative approach is to fit the correlation
function with a damped cosine which approximates the correlation
function calculated using the dispersion relation. Then either the
fitted frequency and damping or values of the interfacial
properties which are consistent with the values of frequency and
damping can be quoted \cite{EM88}. Since there are three
interfacial properties and only two parameters it is necessary, in
this latter case, to make some assumptions.

\section{Results and Discussion}
\subsection{Surface pressure ($\Pi$) - surface concentration ($\Gamma$)
isotherms}
\begin{figure}[h!]
         \begin{center}
          \epsfig{file=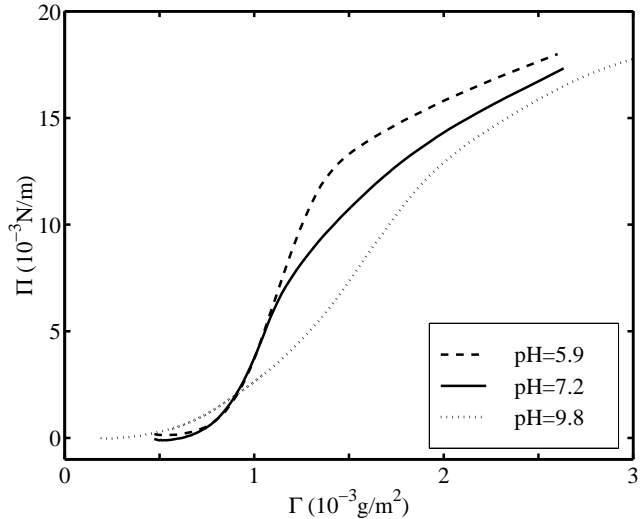,width=8.5cm}
         \end{center}
\label{F5} \caption{Surface pressure ($\Pi$) -- surface
concentration ($\Gamma$) isotherms, obtained using a Wilhelmy
plate, for different pH values, at constant ionic strength
I=0.01M}
\end{figure}

\begin{figure}[h!]
          \begin{center}
         \epsfig{file=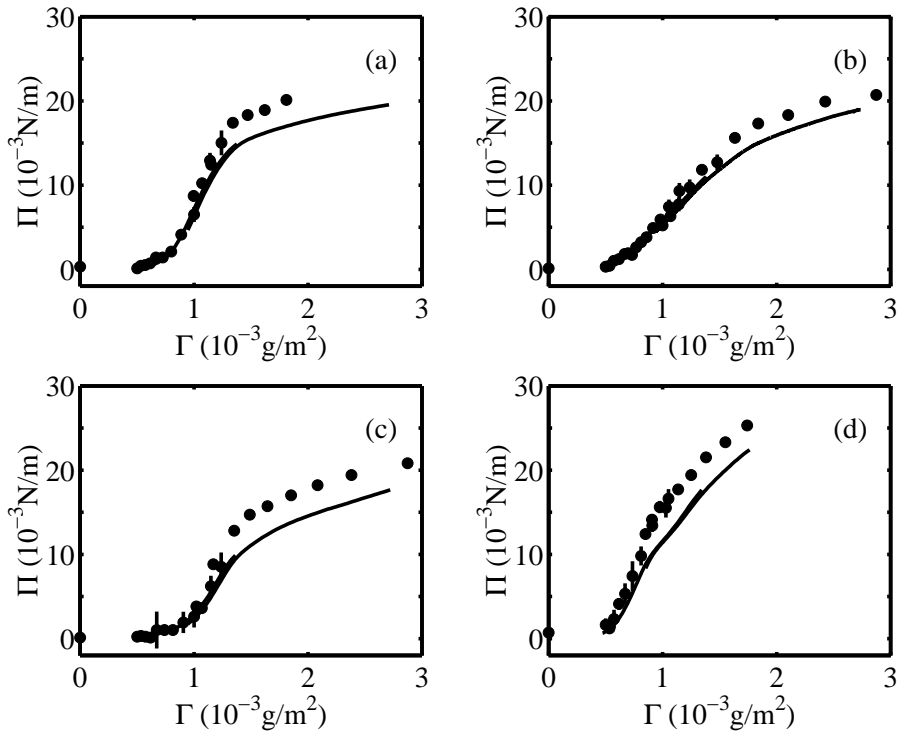,width=12cm}
        \end{center}
\label{F6} \caption{Comparison of `static' $\Pi$--$\Gamma$
isotherms (solid line) to isotherms obtained from surface light
scattering  for different subphase conditions  (points): (a):
pH=5.24, I=0.01; (b): pH=8.30, I=0.01; (c): pH=8.34, I=0.001; (d):
pH=7.60, I=1.1. The scattering vectors q are in the range
$425<q/cm^{-1}< 507$. Error bars are the standard deviation of the
mean arising from the average of values obtained by fitting 5
correlation functions.}
\end{figure}

 Fig.~\ref{F5} shows a selection of surface pressure ($\Pi$) -
surface concentration ($\Gamma$) isotherms for $\beta$--casein
spread on buffers at a
 range of pH values, measured using a Wilhelmy plate.
 %
 %
 %
 %
  The accuracy in
  measuring $\Pi$ is $±0.02\times10^{-3}$N/m. For a single run the random error in
  the surface concentration is negligible, comparing between
  isotherms from different runs
  there is a random error of around 15\% in surface concentration.
  However, once this offset in surface concentration is corrected
   for, the isotherms are highly repeatable.\\ At low concentrations
   the pressure is very low, corresponding to a dilute regime of
   isolated proteins at the surface \cite{JR99}. As the surface
   concentration increases the pressure starts to increase markedly,
   this is the point where the proteins at the surface come into contact
    with each other and marks the onset of the semi--dilute regime $\Gamma=\Gamma^*$. The
    behaviour in this regime clearly varies with the pH, with the
    isotherm becoming flatter as the pH, and thus overall charge,
    is increased. At the highest surface concentrations the isotherms
    appear to converge to a universal concentrated regime, where
    there is significant chain overlap.  We can estimate the  radius of a protein molecule on
the surface from the upturn concentration $\Gamma^*$. $\Gamma$
ranges from 0.25$\times 10^{-3}g/m^2$  for high pH to~0.5$\times
10^{-3}g/m^2$  at the isoelectric pH. This leads to protein
radiuses that increase from 51$\AA$ at the isoelectric pH to
72$\AA$ for high pH (due to the uncertainty in the spreading
procedure the confidence in this result is  $\pm 10 \%$).\\
Fig.~\ref{F6} shows a comparison of $\Pi$--$\Gamma$ isotherms
obtained using Wilhelmy plate methods and surface quasi-elastic
light scattering.
%
%
%
%
The error bars shown
for the SQELS data are the error in the mean for fitting groups of
five correlation functions as described above. At low surface
concentrations there is good agreement between the static and
dynamic measurements. However at higher surface concentrations the
value of the surface pressure measured using SQELS lies around
2$\times10^{-3}$N/m above the static value. This discrepancy is
considerably larger than the uncertainty in the fitting of the
surface pressure. The onset of the deviation always occurs close
to the maximum in the dilational modulus, as discussed below.\\

\subsection{Scaling exponent and the semi-dilute regime}
\begin{figure}[h!]
          \begin{center}
           \epsfig{file=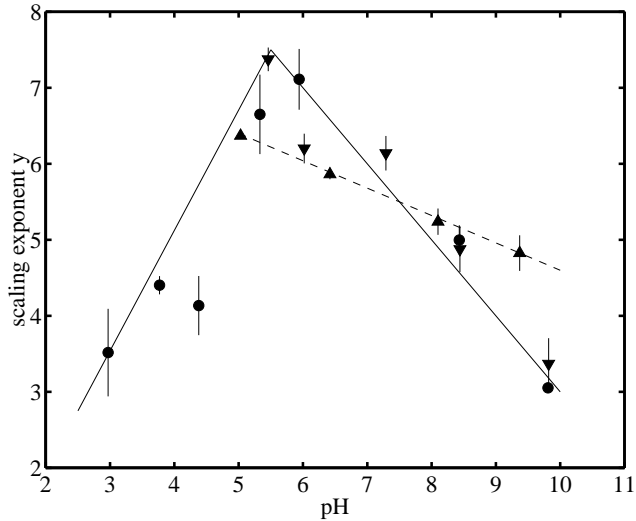,width=8.5cm}
          \end{center}
\label{F7} \caption{Scaling exponent y as a function of pH, for
buffers with different ionic strengths: ($\triangle$):
$0.5<I<1.1$; ($o$): $0.008<I<0.012$; ($\nabla$): $0.001<I<0.003$.
Continuous lines are a guide to the eye for low salt concentration
behavior, the dashed line for high salt. }
\end{figure}

\begin{figure}[h!]
         \begin{center}
           \epsfig{file=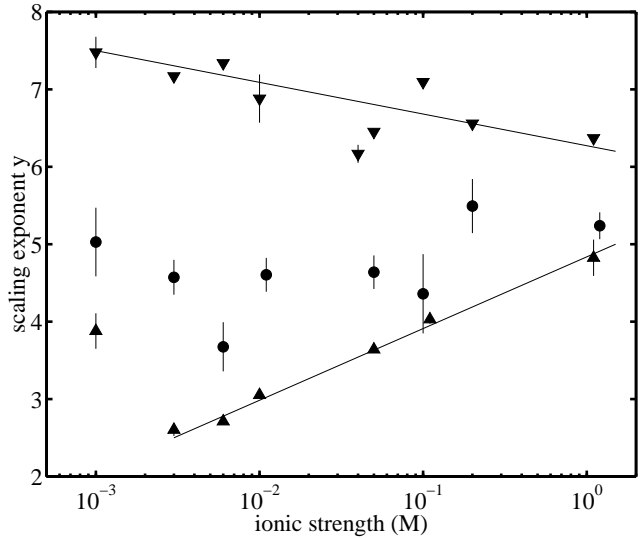,width=8.5cm}
          \end{center}
\label{F8} \caption{Scaling exponent y as a function of ionic
strength, for buffers with different pH: ($\triangle$): $9<pH<10$;
($o$): $7.5<pH<8.5$; ($\nabla$): $5<pH<6$. Lines are a guide to
the eye.}
\end{figure}

 Fig. \ref{F7} shows the variation of the
pressure--concentration scaling exponent, y, as a function of pH.
Data from experiments with three different ionic strengths (I) are
included in this figure.
%
%
%
%
It can be seen that the scaling exponent has a maximum at the
isoelectric point, this indicates that the protein is most compact
at this point. The peak value suggests that $\beta$--casein is
close to $\theta$ condition and agrees with very recent work on
the conformation of polyampholytes \cite{Yam00}. As the pH is
varied away from the isoelectric point the exponent decreases,
approximately linearly with the pH. This change in the exponent
corresponds to a swelling of the polypeptide. The decrease in y is
more dramatic at low I, indicating that screening charge
interactions decreases the swelling effect of a net charge on the
molecule. The exponent y  changes more rapidly on the low pH side
of the isoelectric point, as expected, since the calculated net
charge changes more rapidly here than it does on the high pH side
of the isoelectric point. Where indicated, error bars are the
standard deviation of the mean of 2-8 repeated compressions, and
they indicate the confidence to be expected for this kind of
measure.\\ This behaviour is in contrast to that observed in 3D
polyampholyte gels, where there is a sharp step in the gel
swelling as the overall charge is increased from neutrality
through the threshold charge of eN$^{1/2}$. Here we see a more
gradual change in the degree of swelling. In our system this
critical value of net charge is anticipated to be ca.~3e. This
degree of charging is achieved at a pH very close to the
isoelectric point, therefore it is possible that the highly
collapsed state has been missed.  Also, the lower dimension of the
surface layer compared to the bulk gel may play a part in changing
the nature of the transition from collapsed to swollen states.
This dimensional effect has been recently addressed in a Monte
Carlo simulation of a diblock polyampholyte \cite{Imb99}, where it
is argued that topological constraints modify the 3D coil--globule
transition into a folding in 2D.\\ Fig. \ref{F8} shows the
variation of the scaling exponent with ionic strength for three
ranges of pH.\\
%
%
%
%
For pH5-6, the scaling exponent decreases slightly with increasing
ionic strength, corresponding to a small swelling of the protein.
This behaviour is consistent with that of a polyampholyte, where
the screening out of opposite charge interactions will lead to
chain expansion. An alternative explanation for this small effect
is that increased ionic strength facilitates a small increase in
overall charge leading to an overall chain expansion due to the
increased importance of like charge repulsions.\\ The exponent y
is  constant with I for pH7.5-8.5, meaning that the polyampholyte
tendency to contract is balanced by the net charge present on the
molecule at that pH.\\ For pH9-10, the scaling exponent varies
considerably with ionic strength. It exhibits a minimum at around
I=0.003M, i.e. below this value of I it decreases with increasing
ionic strength and above it it increases. This minimum in the
scaling exponent corresponds to a maximum in the chain swelling,
and the non monotonic behaviour would appear  consistent with
polyampholyte behaviour at low ionic strength followed by
polyelectrolyte behaviour at higher ionic strength. This
conclusion is probably incorrect though, because at the same ionic
strength as the minimum in the scaling exponent (I=0.003M) we find
gross changes in the dilational modulus isotherms. Above this
ionic strength there are two peaks in the dilational modulus and
below there is only one. As discussed in detail below, these
features probably correspond to gross changes in the surface
configuration of the protein, with parts of the molecule moving
into the subphase.  It is therefore possible that the minimum in
the scaling exponent arises from this gross structural change
rather than  a change in behaviour of an essentially homogenous
series of protein configurations. The distribution of positive and
negative charges in $\beta$--casein is non-random, it may  be that
some of the features seen are due to this non-randomness.
Theoretical work \cite{Hig91} suggests that correlations in the
sequence of positive and negative charges are important when the
correlation length is of a similar size to the estimated charged
blob. For the opposite extreme, where the charge sign alternates,
the effect is unimportant. The value of the scaling exponent at
high pH and low salt is in fair agreement with the prediction that
a polyelectrolyte will behave as an extended chain \cite{Joa00}
\cite{Yam00}.\\ Our experimental results  can also be compared to
previous results on the swelling of 3D polampholyte gels as a
function of ionic strength, where non--monotonic swelling behavior
has been observed as a function of ionic strength. English et al
\cite{Eng96} found a maximum in gel swelling, Nisato et al
\cite{NMC99} see only a minimum in swelling as a function of
increasing ionic strength.

\subsection{Dilational modulus: $\epsilon$--$\Gamma$ isotherms}
\begin{figure}[h!]
          \begin{center}
           \epsfig{file=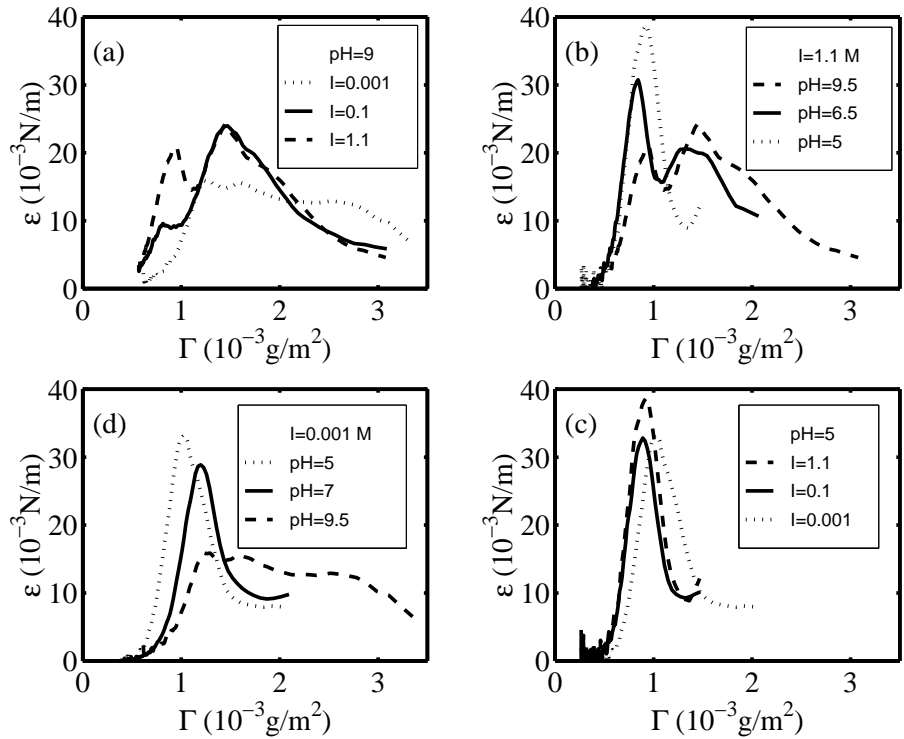,width=12cm}
          \end{center}
\label{F9} \caption{Dilational modulus $\epsilon_{st}$ obtained
from surface pressure isotherms over a range of subphase
conditions. (a) and (c) show the effect of varying the ionic
strength I at fixed pH, (b) and (d) the effect of varying pH at
fixed I.}
\end{figure}

 The
scaling exponents in the previous section probe the proteins when
they have just started to come into contact with one another. We
will now use the dilational modulus ($\epsilon$) to probe the
behaviour as the protein is further compressed. Dilational
moduli--surface concentration plots for a range of pH and ionic
strength are shown in Fig.~\ref{F9}.
%
%
%
%
 Each figure
contains data where one of pH and ionic strength is held constant
whilst the other is varied. (a) shows data for fixed pH = 9.5 and
ionic strength varying from 0.001 to 1.1, here two peaks are seen
in $\epsilon$, one at around $\Gamma=0.8\times10^{-3}$g/m$^2$ and
the other at around 1.5$\times10^{-3}$g/m$^2$, the magnitude of
both of these two peaks increases as a function of increasing
ionic strength. At the lowest ionic strength the peak at lower
$\Gamma$ is not apparent, (b) shows data for fixed ionic
strength=1.1 and pH varying from 5 to 9.5. Once again two peaks
are observed in the dilational modulus at $\Gamma$ = 0.8 and
1.5$\times10^{-3}$g/m$^2$, although they shift to slightly higher
$\Gamma$ as pH is reduced. In contrast to the data in (a), the
magnitude of the peak at $\Gamma$ = 0.8 increases with reducing pH
but that at higher $\Gamma$ decreases in magnitude. (c) shows data
for fixed pH=5.0 and ionic strength varying between 0.001 and 1.1,
here only a single peak in the dilational modulus is observed
which shifts to slightly higher $\Gamma$ as ionic strength is
reduced and shows little change in magnitude. Finally, (d) shows
data at fixed ionic strength = 0.001 and pH varying from 5 to 9.5
once again only one peak in the dilational modulus is observed,
this shifts to higher $\Gamma$ with increasing pH and reduces in
intensity.\\ Note that increasing ionic strength corresponds to
reducing the electrostatic screening length from 96$\AA$ to
3$\AA$. Increasing pH above 5 leads to increased overall charge on
the protein.\\

\begin{figure}[h!]
          \begin{center}
           \epsfig{file=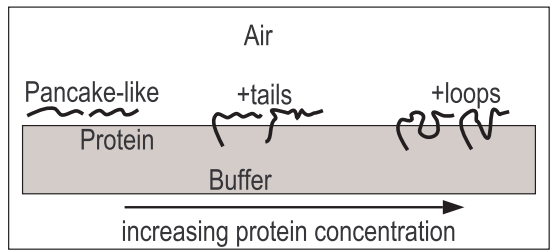,width=8.5cm}
          \end{center}
\label{F10} \caption{Schematic diagram of the $\beta$--casein
molecule at the air/buffer interface. At low concentration
$\Gamma$ the molecule lies on the surface with a `pancake'
configuration. As $\Gamma$ increases, first tails then loops of
molecule are forced into the subphase.}
\end{figure}

Previously, for $\beta$--casein, only a single peak in the
dilational modulus at lower $\Gamma$ has been commented upon,
being attributed either to the collapse of the N-terminal end of
the protein into the subphase \cite{Mel98}, or to looping of parts
of the molecule in the subphase \cite{NSP99}. We propose that the
first peak corresponds to the tail collapse in the subphase, and
that the second peak arises from the collapse of a second region
of the molecule into the subphase, this may well be the loop
region identified by examination of the primary sequence discussed
above. These transitions that depend on concentration,  pH and I
are illustrated schematically in Fig.~\ref{F10}.\\
%
%
%
%
To our knowledge no synthetic system, even a multiblock copolymer,
has been shown to exhibit two maxima in the dilational modulus as
a function of surface concentration.

\subsection{Dilational modulus: Surface Transitions}
\begin{figure}[h!]
          \begin{center}
           \epsfig{file=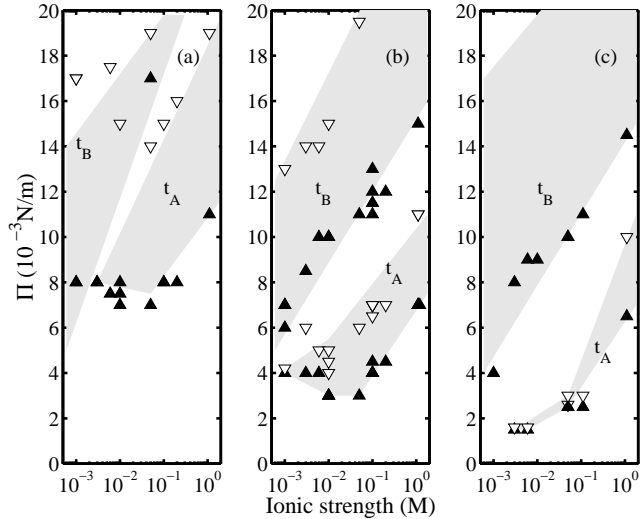,width=8.5cm}
          \end{center}
\label{F11} \caption{`Configuration maps' at fixed pH and varying
I, obtained from $\epsilon_{st}$--$\Pi$ isotherms. ($\triangle$)
are the location of  maxima in $\epsilon_{st}$--$\Pi$ plots and
($\nabla$) correspond to minima. Shaded areas, labelled t$_A$ and
t$_B$, guide the eye to regions of decreasing $\epsilon$ with
increasing $\Pi$. We tentatively associate t$_A$ as the region
where the molecule forms tails in the subphase, and  t$_B$ as
where it additionally forms loops. (a): $5<pH<6$; (b):
$7.5<pH<8.5$; (c): $9<pH<10$}
\end{figure}

\begin{figure}[h!]
          \begin{center}
           \epsfig{file=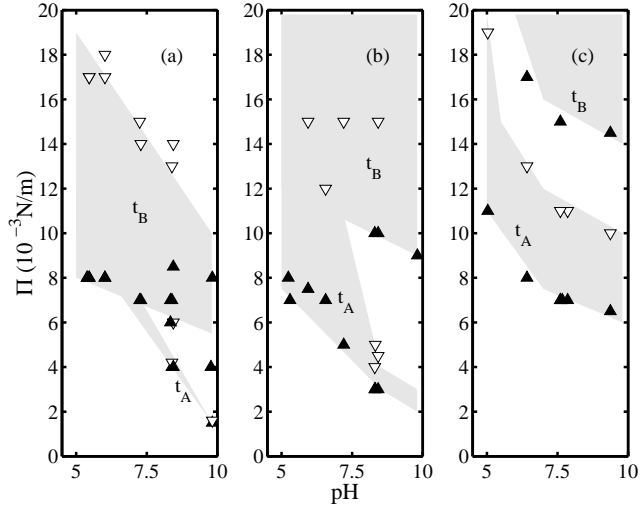,width=8.5cm}
          \end{center}
\label{F12} \caption{`Configuration maps' at fixed I and varying
pH, obtained from $\epsilon_{st}$--$\Pi$ isotherms. ($\triangle$)
are the location of  maxima in $\epsilon_{st}$--$\Pi$ plots and
($\nabla$) correspond to minima. Shaded areas, labelled t$_A$ and
t$_B$, guide the eye to regions corresponding to a decreasing
$\epsilon$ with increasing $\Pi$. We tentatively associate t$_A$
as the region where the molecule forms tails in the subphase, and
t$_B$ as where it additionally forms loops.  (a): $0.001<I<0.003$;
(b): $0.008<I<0.012$; (c): $0.5<I<1.1$}
\end{figure}

We have further analysed this data by observing the pressures
where maxima and minima in the dilational modulus occur, as a
function of pH and ionic strength (I). By considering a maximum in
$\epsilon$ as the onset of a conformation transition and the
successive minimum as the end of the transition, we are in the
position to present ``configuration maps'' of the protein
monolayer.\\ In Fig.~\ref{F11} we show the effect of ionic
strength.
%
%
%
%
Shaded regions are for surface concentrations where $\epsilon$ is
decreasing, these are the regions in which transitions between
configurations occur. The transition corresponding to the first
peak in $\epsilon$ is labeled as t$_A$ (we are proposing that this
is the tail protruding in the subphase), and the transition
corresponding to the second peak is t$_B$ (this would be the
looping of some segments). The most obvious effect of salt
addition is to move all transitions to higher $\Pi$. This means
that transitions become more energetically costly the more charge
interactions are screened. From Fig.~\ref{F11} it is now clear how
a two peak structure turns into a single peak for very low I. At
the isoelectric pH, Fig.~\ref{F11}~(a), one sees that the
transition t$_A$ does not occur at low I. This is consistent with
the picture that a neutral polyampholyte is compact and tightly
folded. Thus there no tail can easily be submerged. On the
contrary at the other limit of high pH in  Fig.~\ref{F11}~(c), the
tail transition t$_A$ appears well defined and at low I it
disappears at $\Pi=0$, meaning that the tail is already submerged
at $\Gamma=\Gamma^*$ when the polymers first come into contact.\\
In Fig.~\ref{F12} we show the effect of pH.
%
%
%
%
The general feature is that as the pH increases and the molecule
develops a net charge, all transitions occur at lower surface
pressure.\\ In Fig.~\ref{F12}~(a), low~I, one sees  that the tail
transition t$_A$ is not resolved from t$_B$ at pH5, and emerges
from pH around 7.5. It is very thin, meaning that the energy
difference between conformations with the tail submerged or on the
surface is very small. Also it appears that t$_A$ is moving to
$\Pi=0$, thus confirming that at high pH and low I we expect the
tail to be spontaneously submerged.\\ Moving to Fig.~\ref{F12}~(b)
and~(c) one sees that t$_A$ broadens with increasing I, and the
loop transition t$_B$ shifts towards the upper $\Pi$ limit of our
study.\\ We note that previous studies \cite{NSP99} have shown the
$\beta$--casein monolayer to wholly collapse into a multilayer at
pressures around 22$\times10^{-3}$N/m, so for
$\Pi\geq20\times10^{-3}$N/m it would not be very significant to
explain $\epsilon$ behaviour in terms of single molecule
conformation.\\

The concentrations at which the peaks in the dilational modulus
occur are consistent with the calculated area covered by  the
protein with, successively, a terminal tail and a loop in the
subphase. The peaks shift to lower concentrations with increased
pH, this is consistent with  the protein extending as more groups
dissociate. The lower concentration peak reduces in magnitude as
pH increases, whilst the higher concentration peak increases in
magnitude with increasing pH. This is not unexpected, the free
energy cost of forcing a longer loop into the subphase increases
with loop length, since the ends of the loop must be constrained
to the interface. The entropic cost of tail submersion depends
less on length, as only one end is held at the surface.

\subsection{Dilational modulus: High Frequency}
\begin{figure}[h!]
          \begin{center}
           \epsfig{file=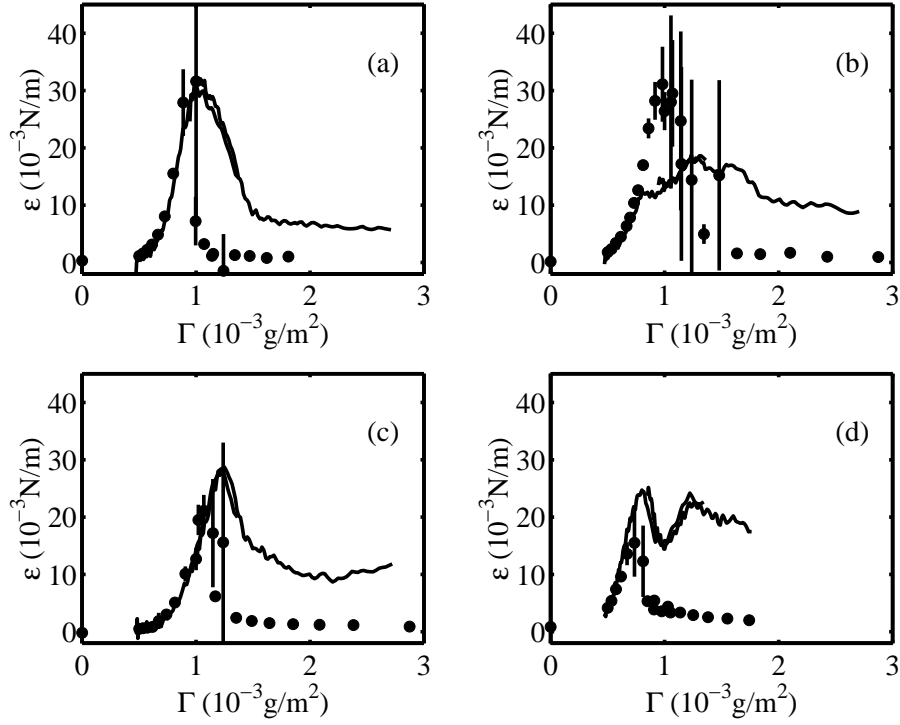,width=12cm}
          \end{center}
\label{F13} \caption{Comparison of `static' $\epsilon$--$\Gamma$
isotherms (solid line) to dilational moduli obtained from surface
light scattering for different subphase conditions (points): (a):
pH=5.24, I=0.01; (b): pH=8.30, I=0.01; (c): pH=8.34, I=0.001; (d):
pH=7.60, I=1.1;}
\end{figure}

Fig. \ref{F13} shows dilational moduli, obtained using SQELS as a
function of surface concentration, compared with the values
obtained from the static $\Pi$--$\Gamma$ isotherm.
%
%
%
%
At low surface concentration the static and dynamic values of
$\epsilon$ are very similar, in common with the surface pressure
measurements. At higher surface concentrations the dynamic,
$\epsilon$, value lies considerably below the static value. Once
again the divergence of static and dynamic results occurs at or
shortly before the first major maximum in the dilational modulus.
The dynamic value of the dilational modulus only ever contains one
strong maximum, unlike the static value which sometimes exhibits a
peak at higher surface concentration. The significantly reduced
value of the dynamic dilational modulus as compared to the static
value has often been observed with soluble surfactants
\cite{Lan92}. In these systems the difference arises from
diffusion of surfactant out of the surface layer during the period
of the dilational waves at the surface. This has been described
theoretically by Lucassen et al \cite{LT72} and Hennenberg et al
\cite{Hen92}. Here we do not anticipate a gross diffusional motion
of the protein into and out of the surface layer at such high
surface concentrations. However it is probable that the motion of
parts of the molecule into and out of the surface layer would have
a similar signature \cite{MOR99}. Lucassen et al find, (here we
use the notation of Langevin):
\begin{eqnarray}
\epsilon_0\,=\,\epsilon_{st}\,\left(\,\frac{1\,+\,\Omega}{1\,+2\,\Omega\,+\,2\,\Omega^2}\,
\right)\,+\,i\epsilon_{st}\,\left(\,\frac{\Omega}{1\,+2\,\Omega\,+\,2\,\Omega^2}\,
\right)
 \label{eq12}
\end{eqnarray}
where $\Omega$ is the reduced frequency,
\begin{eqnarray}
 \Omega\,=\,\sqrt{\frac{D}{2\,\omega}}\,\frac{dc}{d\Gamma}\,=\,\sqrt{\frac{1}{\omega\,\tau_c}},
 \label{eq13}
\end{eqnarray}
D is the diffusion coefficient of the surface species and c is the
bulk concentration of adsorbant and thus $\tau_c$ is the
characteristic time for this process of segment exchange between
surface and subphase.

\subsection{Dilational modulus: SQELS as a function of q at high surface concentration}
\begin{figure}[h!]
          \begin{center}
           \epsfig{file=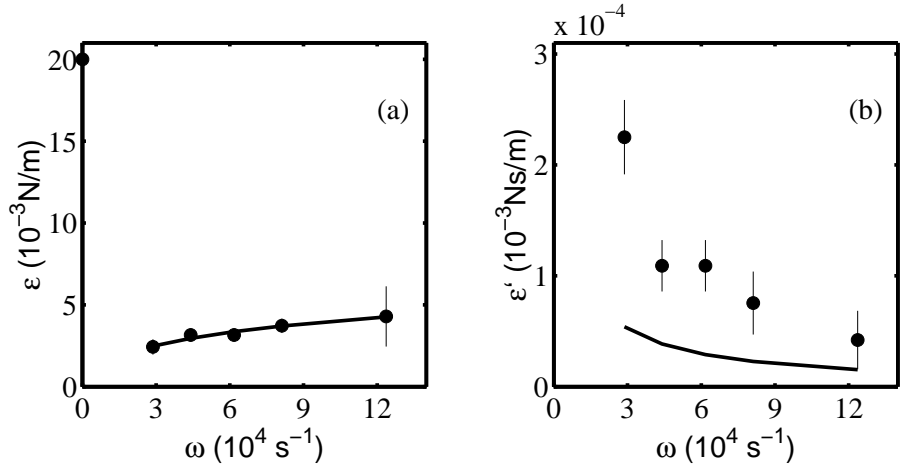,width=12cm}
          \end{center}
\label{F14} \caption{Frequency dependence of the dilational
modulus (a) and dilational viscosity (b) at pH=7.6, I=1.1,
$\Pi=12\times 10^{-3}N/m$. This pressure is in the high
concentration region, where the high frequency $\epsilon$ is
consistently lower than the static $\epsilon_{st}$. Solid lines
both in (a) and (b) are the result of fitting the data in (a) with
the model described by eq.~(\ref{eq12}). In~(a) the $\epsilon$
value plotted at $\omega=0$ is $\epsilon_{st}$ from the
compression isotherm.}
\end{figure}

By varying the scattering vector at which we collect SQELS data we
vary the frequency, $\omega$, of the capillary waves we observe.
This data can then be tested against (\ref{eq12}). This is done in
Fig.~\ref{F14}.
%
%
%
%
 The data are in the high surface concentration regime
where the static dilational modulus is larger than the SQELS
modulus. A fit to the real part of the dilational modulus was made
by varying the parameter $\tau_c$ and the static dilational
modulus, $\epsilon_{st}$. Fitted values are
$\epsilon=9.2\times10^{-3}$N/m and $\tau_c=12 \mu$s. The value of
$\epsilon_{st}$ from the compression isotherm is
20$\times10^{-3}$N/m. If we estimate
  the value of dc/d$\Gamma$
from data presented by Graham and Philips \cite{GP79} the value of
D obtained in this way is $7.5\times10^{-11}m^2/s$. The Lucassen
model produces values of the dilational elasticity and viscosity
of approximately the right magnitude, with the correct dependence
on frequency, however our data show a value of the complex part
($\omega\times \epsilon' $) which is larger than the real part
($\epsilon$) and this is cannot be accounted for by (\ref{eq12}).
The discrepancy may be due either to a systematic error in the
determination of the dilational viscosity or to the fact that the
type of behaviour we anticipate is simply not modelled well by
these expressions. Since the proposed mechanism is not simply
diffusion into the bulk, the fitted values of $\tau_c$ might not
have a real physical significance.

\begin{figure}[h!]
          \begin{center}
           \epsfig{file=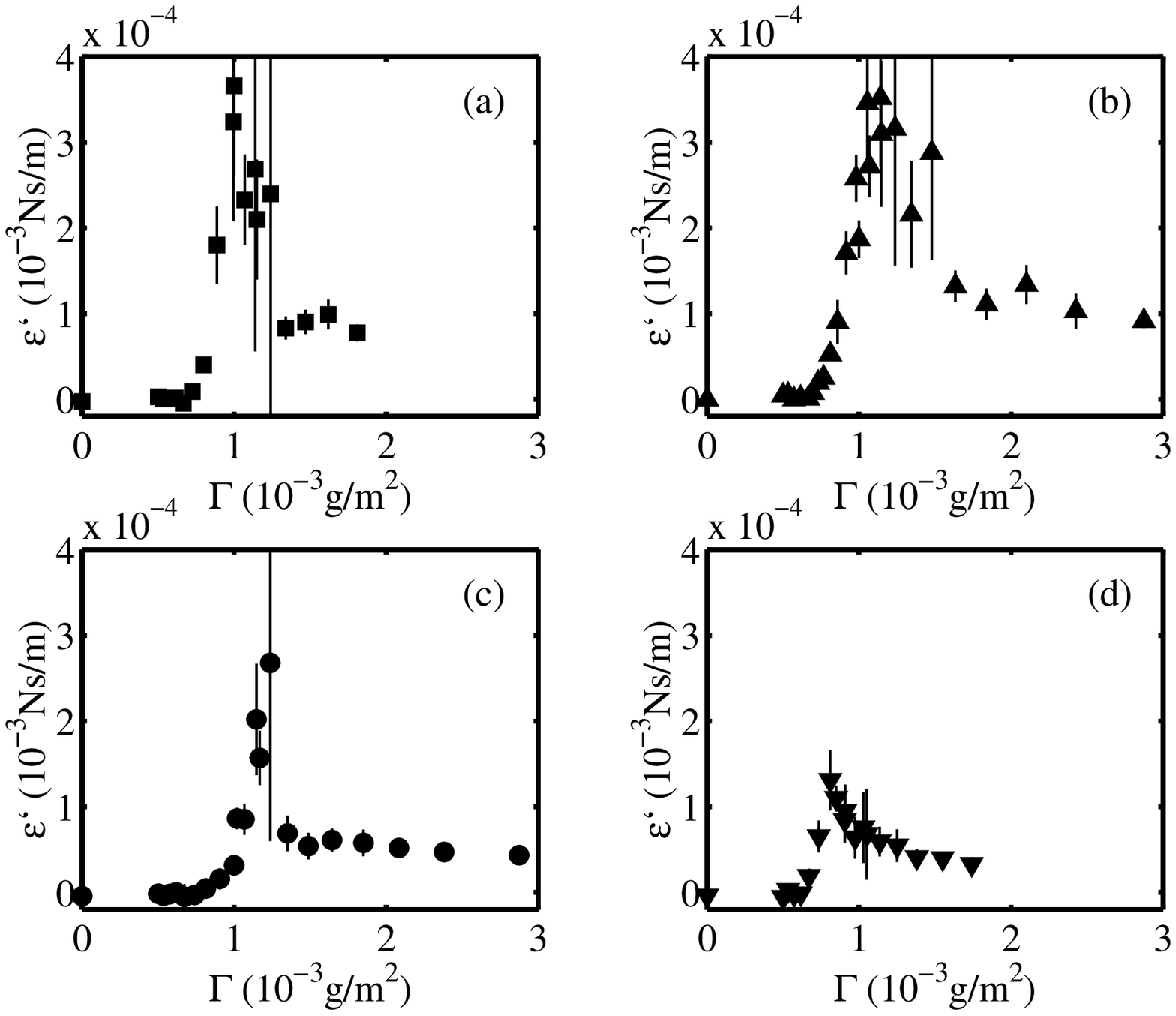,width=12cm}
          \end{center}
\label{F15} \caption{ Dilational viscosities obtained from surface
light scattering for different subphase conditions: (a): pH=5.24,
I=0.01; (b): pH=8.30, I=0.01; (c): pH=8.34, I=0.001; (d): pH=7.60,
I=1.1;}
\end{figure}

Fig.~\ref{F15} shows values of the dilational viscosity obtained
using SQELS. We have not measured  low frequency values with which
to compare this data. However we find that the dilational
viscosity of $\beta$--casein is very similar to that reported in
the literature for other polymer monolayers  measured by SQELS
\cite{MOR99}.
%
%
%
%
The dilational viscosity exhibits a maximum at the same surface
concentration as the maximum in the dynamic dilational modulus.

\subsection{SQELS as a function of q at low surface concentrations} 
\begin{figure}[h!]
          \begin{center}
           \epsfig{file=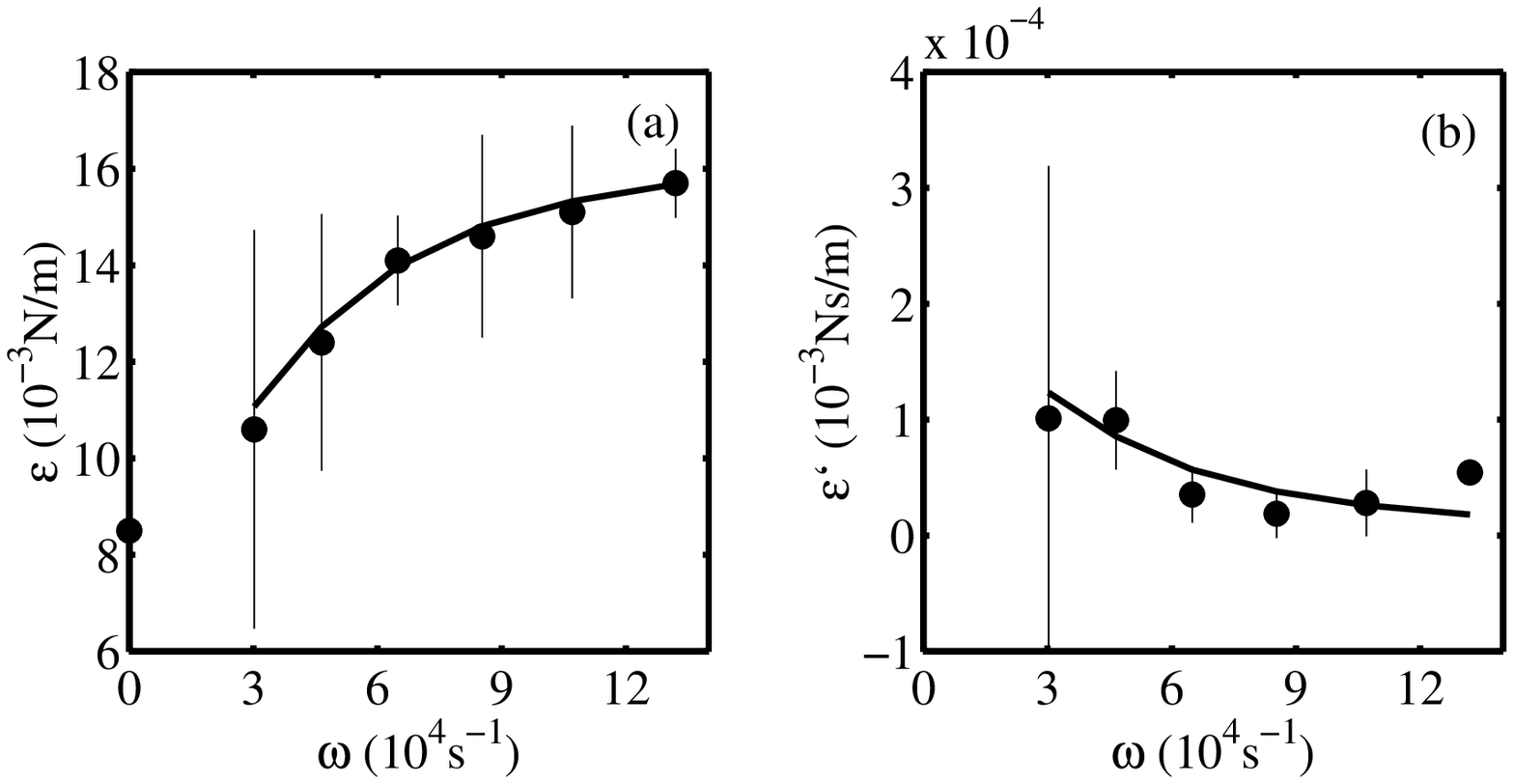,width=12cm}
          \end{center}
\label{F16} \caption{Frequency dependence of the dilational
modulus (a) and dilational viscosity (b) at pH=8.3, I=0.01,
$\Pi=2\times 10^{-3}N/m$.  This pressure is close to the first
conformation transition region, where, for some subphase
conditions, we observe a high frequency $\epsilon$ bigger than the
static $\epsilon_{st}$. Solid lines are best fits with the model
described by eq.~(\ref{eq14}). In~(a) the $\epsilon$ value plotted
at $\omega=0$ is $\epsilon_{st}$ from the compression isotherm.}
\end{figure}

In the
previous section data from a range of q vectors (and thus
frequencies) were introduced for layers at high surface
concentrations. Here we consider similar data acquired at low
surface concentrations. Fig.~\ref{F16} shows dilational modulus
and viscosity as a function of q for a protein with surface
concentration $0.8\times10^{-3}$g/m$^2$ on a buffer with pH=8.3
and I=0.01M.
%
%
%
%
 This corresponds to a point close
in concentration to the first peak in the dilational modulus,
where the dynamic values of dilational modulus lie slightly above
the static values (see Fig.~\ref{F13}~(d)). We fit these data
using a Maxwell fluid model, i.e. a serial combination of a spring
and dashpot, characterised by a single relaxation time, $\tau_m$.
The complex dilational modulus is given by:
\begin{eqnarray}
\epsilon_0\,=\,\left(\,\epsilon_{st}\,+\,\epsilon_{\infty}\frac{(\omega\,\tau_D)^2}{1\,+\,(\omega\,\tau_D)^2}\,
\right)\,+\,i\,\left(\,\epsilon_{\infty}\frac{\omega\,\tau_D}{1\,+\,(\omega\,\tau_D)^2}\,
\right)
 \label{eq14}
\end{eqnarray}
where the first term is the dilational (elastic) modulus and the
second part is the dilational viscosity. $\epsilon_{st}$ is the
static dilational modulus, $\epsilon_{\infty}$ is the amplitude of
the relaxation and $\omega$ is the frequency of measurement. The
fit to the data is shown by a solid line, a single relaxation time
of 23$\mu$s. We believe that this relaxation time is related to a
tail submersion event during the period of the dilational wave. It
only appears strongly at high pH and moderately low I, that is,
 as discussed above (see Fig.~(\ref{F11}) and~(\ref{F12})), the conditions
where the energy difference between absorbing and desorbing the
tail is small. Assuming, as we have found, that an expression such
as (\ref{eq14}) holds, then the relaxation time $\tau_D$ can be
obtained from the static dilational modulus and the dynamic
dilational modulus and viscosity measured at a single frequency:
\begin{eqnarray}
 \tau_D\,=\,\frac{\epsilon_0\,-\,\epsilon_{st}}{\omega\,\epsilon '}
 \label{eq15}
\end{eqnarray}
The results of this analysis (data not shown) are consistent with
the more complete frequency measurements. Such relaxation times
have been measured using SQELS by a number of groups for a number
of polymer systems. Monroy et al \cite{MOR98} found an Arrehnius
temperature of the relaxation time for polyvinylactetate (PVAc)
(M$_w$ = 90000), at the overlap concentration $\Gamma^*$. The
relaxation time decreased from 80$\mu$s at 1$^\circ$C to 5$\mu$s
at 25$^\circ$C. Monroy also finds relaxation times for mixed
monolayers of PVAc and poly(4-hydroxystyrene) (P4HS) \cite{MOR99}.
The relaxation times are shorter for higher surface coverages and
for larger fractions of the PVAc. Brown et al \cite{Bro99} find
relaxations times of this magnitude which reduce as a function of
increasing surface concentration for polymethylmethacyrlate -
poly-4-vinyl pyridine diblock copolymers. Richards et al
\cite{RRT96} find decreasing values of the relaxation time at
lower concentrations, for PMMA-polyethylene oxide diblock
copolymers, but at higher concentrations the relaxation times
increase with increasing concentration. Mizuno et al \cite{Miz00}
also find a Maxwell relaxation, and they note that its physical
origin can be a potential barrier to desorption.

\subsection{Pressure relaxation data}
\begin{figure}[h!]
          \begin{center}
           \epsfig{file=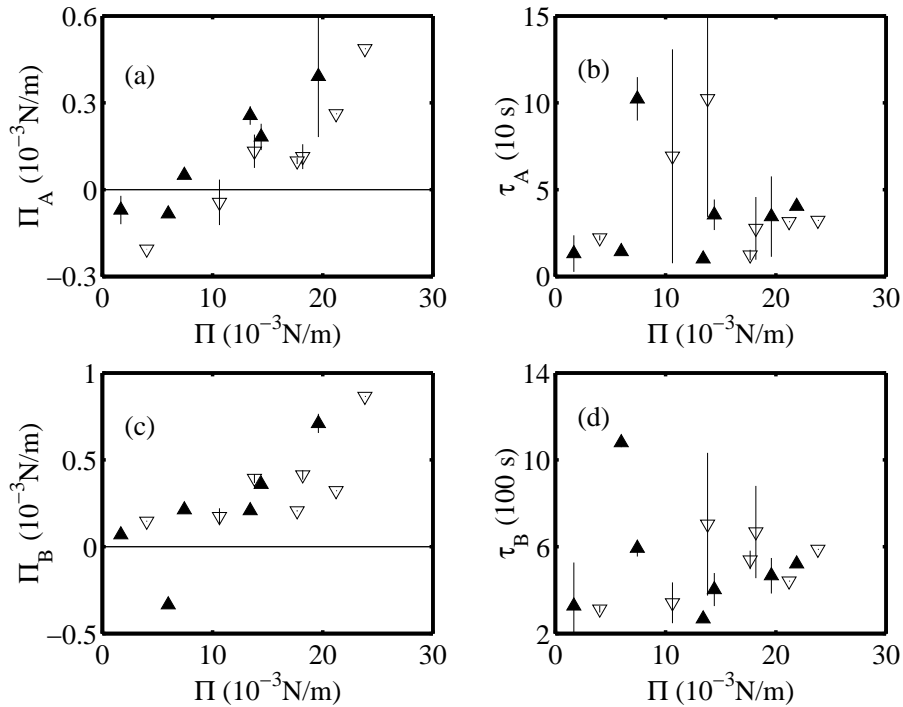,width=12cm}
          \end{center}
\label{F17} \caption{Parameters describing the relaxation of
pressure after a `step--compression', fitted with the form shown
in eq.~(\ref{eq11}). ($\triangle$) correspond to a buffer pH=5.60,
I=0.1 and ($\nabla$) to pH=8.32, I=0.01}
\end{figure}

  We have carried out `step compressions', where the surface
layer is compressed at a constant rate for a short time and the
evolution of the pressure ($\Pi$) is then measured for about 15
minutes, following which the next compression is performed.
Similar measurements on other systems are reported for example by
Monroy et al \cite{MOR98}. Such measurements have an obvious
parallel with step strain measurements made on bulk materials.\\
We find  that a single exponential decay is insufficient to fully
describe the data (this was also observed in \cite{NSP99} at high
$\Pi$), but that in every case the spectra can be fitted with a
sum of two exponentials with  well separated time constants:
\begin{eqnarray}
\Pi(t)\,=\,\Pi_A\,e^{-t/\tau_A}\,+\,\Pi_B\,e^{-t/\tau_B}\,+\,\Pi_C
 \label{eq11}
\end{eqnarray}
where $\Pi_A$ and $\Pi_B$ are the amplitudes of relaxations with
time constants $\tau_A$ and $\tau_B$. We find that the fast mode
timescale is $20<\tau_A/s<50$ and the slow mode  timescale is
$300<\tau_B/s<700$. $\Pi_C$ is a constant pressure offset (the
equilibrium pressure) and t is the time since the end of the last
constant rate compression. Fig.~\ref{F17} shows fitted values for
these parameters from two sets of step compressions, measured on
buffers at the isoelectric pH with I=0.1 and at pH=8.3 with
I=0.01. These buffer conditions are chosen because they correspond
to very different isotherms.  The ratio between successive areas
is always $\Delta A/A=0.55$.
%
%
%
%
We are exploring a more complete range of subphase conditions and
we refer to a future work for a complete presentation of these
results \cite{CH01}.\\ Fig.~\ref{F17}~(a) and~(c) show the
amplitudes $\Pi_A$ and $\Pi_B$ respectively of the fast and slow
relaxations, as a function of the equilibrium pressure $\Pi_C$.
One sees a general trend for the relaxations to become bigger as
the pressure increases. At low $\Pi$ negative values of both
$\Pi_A$ and $\Pi_B$ are sometimes found. These are puzzling, as a
negative amplitude reflects the fact that the pressure increases
after the compression has ended. Although the data presented is
too limited to support a strong conclusion, we observe that
negative amplitudes for the slow mode ($\Pi_B$) occur only for
buffer conditions that present a ``tail submersion'' transition,
and then only around the transition pressure.\\ We believe that
the fast mode is connected to relaxation of stress in the
direction parallel to the compression, and that the slow mode is
connected to stress relaxation between the parallel and normal
directions. These stresses arise because the fast step compression
induces a concentration gradient parallel to the compression, and
can also bring the monolayer to a transient state where the strain
in the direction of compression is greater than in the normal
direction. We further suggest that below the ``loop transition''
pressure (which is in most conditions around $\Pi=10\times
10^{-3}$N/m, see Fig.~\ref{F12}) the surface pressure relaxations
can be understood on the basis of the concavity of the equilibrium
$\Pi$--$\Gamma$ isotherm. On the contrary, above the loop
transition, a fast compression does not give time to the monolayer
to equilibrate with the subphase, and thus one can not refer to
the equilibrium isotherm.\\ In Fig.~\ref{F17}~(b) and~(d) we show
the timescales $\tau_A$ and $\tau_B$ respectively of the fast and
slow relaxations, as a function of the equilibrium pressure
$\Pi_C$.\\ Noskov \cite{Nos95} has  modelled these relaxations,
and his work is relevant to understanding  the observed relaxation
timescales. He describes the interfacial relaxations as analogous
to the Rouse-like relaxation modes of a bulk gel and how the
presence of loops and tails of the polymer in the subphase  leads
to some degree of entanglement.\\

\section{Conclusions}
We have measured the interfacial behaviour of $\beta$--casein as a
function of pH and ionic strength with high frequency light
scattering and conventional methods. We find scaling exponents in
the ``semi--dilute'' regime that agree with calculations reported
in the literature for either polyampholytes or polyelectrolytes,
depending on the  charge present on the protein. In particular, we
find that close to overall neutrality the protein molecule behaves
like a random walk; as the pH is moved from the isoelectric pH the
protein expands continuously almost reaching an extended chain
configuration. At high pH and very low salt concentration, where
the protein is overall negative, we observe an effective swelling
then deswelling of the chain as a function of increasing ionic
strength. At higher ionic strength the chain acts like a
polyelectrolyte. Here only short range interactions are possible
and, since the chain is overall charged, these interactions will
be predominantly repulsive, weakening as the screening length is
further reduced thus leading to deswelling of the chain.\\
Interfacial relaxation times have also been measured using both
surface quasi-elastic light scattering and strain relaxation
measurements. These suggest that at low surface concentrations the
proteins act like independent disks, whilst at higher surface
concentrations the layer forms an entangled network with
Rouse-like relaxation modes.\\ The dilational modulus exhibits one
or two peaks as a function of surface concentration, depending on
the buffer conditions. These peaks are associated with the
collapse of hydrophilic parts of the molecule into the subphase
and we present a ``configuration map'' showing the dependance of
these surface transitions on pH and ionic strength. We suggest,
tentatively, that the peak at lower concentrations corresponds to
the penetration of a tail from one end of the molecule and that at
the higher concentration is associated with a loop. This
conclusion is supported by the discovery of prospective loop and
tail regions in the amino acid sequence of the protein, whose
sizes broadly match the positions at which the peaks occur in the
dilational modulus.
\\The protein $\beta$--casein was chosen in part because as a random coil
protein it should most resemble a synthetic or model
polyampholyte. We believe this is the first such study of this
type, in applying ideas relating to polyampholytes to the
understanding of proteins at interfaces. We are currently
conducting a similar study into the behavior of a globular
protein, for comparison to $\beta$--casein. We expect the
relevance  of the observed behavior of $\beta$--casein  to depend
on the extent to which the globular protein  unfolds at the
surface.

\section{Acknowledgements}
 We are grateful to Unilever Plc and EPSRC for
funding and to Prof. Randal Richards and Dr Mark Taylor for advice
and encouragement regarding the design and construction of the
SQELS apparatus. We would also like to thank Dr Peter Wilde, Dr
Martin Buzza and Peter Bermel for useful discussions.

%
%
%
%

%

\clearpage
\begin{thebibliography}{99}
\bibitem{Kud99}    S. E. Kudaibergenov, Adv. Polym. Sci. {\bf{144}}, 120 (1999).

\bibitem{HJ91}   P. G. Higgs and J.-F. Joanny,   J. Chem. Phys.
{\bf{94}}, 1543 (1991).

\bibitem{LT00}   N. Lee and D. Thirumalai,  J. Chem. Phys. {\bf{113}}, 5126 (2000).

\bibitem{KK95}    Y.Kantor and M. Kardar, Phys. Rev. E {\bf{51}}, 1299 (1995).

\bibitem{LO98}   N. Lee and S. Obukhov, European Physical Journal B {\bf{1}}, 371
(1998).

\bibitem{Yam00} V.Yamakov, A.Milchev, H.J.Limbach, B.D\"{u}nweg,
R.Everaers, Phys. Rev. Lett. {\bf{85}}, 4305 (2000).

\bibitem{TT00}   M. Tanaka and T. Tanaka, Phys. Rev. E {\bf{62}}, 3803
(2000).



\bibitem{DR95}   A. V. Dobrynin and M. Rubenstein, J.  Phys. II (France) {\bf{5}},
677 (1995).

\bibitem{NMC99}   G. Nisato, J. P. Munch, and S. J. Candau, Langmuir {\bf{15}},
4236 (1999).

\bibitem{Eng96}   A. E. English, S. Maf\'{e}, J. A. Manzanares, X. Yu, A. Y.
Grosberg, and T. Tanaka,  J. Chem. Phys. {\bf{104}}, 8713 (1996).

\bibitem{CC93}  J.-M. Corpart and F. Candau, Macromolecules {\bf{26}}, 1333
(1993).

\bibitem{JR99} R.A.L.Jones and R.W.Richards, {\em{Polymers at Surfaces and
Interfaces}}
(Cambridge University Press, Cambridge, 1999).

\bibitem{Ada97}  A.W.Adamson and A.P.Gast, {\em{Physical Chemistry of Surfaces}}, Sixth Edition,   (Wiley, New York,
1997).

\bibitem{Mob98} D.M\"{o}bius and R.Miller, {\em{Proteins at Liquid Interfaces} }  (Elsevier, Amsterdam,
1998).


\bibitem{DJ76}   M. Daoud and G. Jannink, J. Phys. (France) {\bf{37}},
973 (1976).

\bibitem{Lan92} D.Langevin, {\em{Light Scattering by Liquid Surfaces and Complementary
Techniques}}
  (Dekker, New York, 1992).

\bibitem{Dic93}  E. Dickinson, D. S. Horne, J. S. Phipps, and R.
M. Richardson, Langmuir {\bf{9}}, 242 (1993).

\bibitem{GP79}  D. E. Graham and M. C. Philips, J. Colloid Int. Sci. {\bf{70}}, 415 (1979).

\bibitem{HKC91}   J. R. Hunter, P. K.
Kilpatrick, and R. G. Carbonell,  J. Colloild Int. Sci.
{\bf{142}}, 429 (1991).

\bibitem{Mel98}  M. Mellema, D. C. Clark, F. A. Husband, and A. R. Mackie,
Langmuir {\bf{14}}, 1753 (1998).

\bibitem{APR00} R. E. Anderson, V. S. Pande, and C. J. Radke,  J. Chem. Phys. {\bf{112}}, 9167 (2000).

\bibitem{Lee96}  F. A. M. Leermakers, P. J. Atkinson, E. Dickinson, and D. S.
Horne,  J. Colloid Int. Sci. {\bf{178}}, 681 (1996).

\bibitem{Dou94}  R. Douillard, M. Daoud, J. Lefebvre, C. Minier, G. Lecannu,
and J. Coutret,  J. Colloid Int. Sci. {\bf{163}}, 277 (1994).

\bibitem{Agu99}   V. Agui\'{e}-B\'{e}ghin, E. Leclerc, M. Daoud, and R.
Douillard,  J. Colloid Int. Sci. {\bf{214}}, 143 (1999).

\bibitem{FM99}  V. B. Fainerman and R. Miller, Langmuir {\bf{15}}, 1812 (1999).

\bibitem{BA00}  A. Bairoch and R. Apweiler, Nucleic Acids Research {\bf{28}},
45 (2000).


\bibitem{Swa82}  H. E. Swaisgood, in {\em{Developments in Dairy Chemistry}}, ed
P. F. Fox (Applied Science Publishers, London, 1982), Vol. 1, pp.
1-59.




\bibitem{Cal95}  P. Calmettes, D. Durand, V. Receveur, M. Desmadril, P. Minard,
and R. Douillard, Physica B {\bf{213}}, 754 (1995).

\bibitem{Daw59} R. M. C. Dawson, D. C. Elliot, W. H. Elliot, and K. M. Jones,
{\em{Data for Biochemical Research}} (Clarendon Press, Oxford,
1959).

\bibitem{NSP99} M. R. R. Ni\~{n}o, C. C. S\'{a}nchez, and J. N. R.
Patino, Coll. and Surf. B: Biointerfaces {\bf{12}}, 161 (1999).



\bibitem{EM87}
J. C. Earnshaw and R. C. McGivern, J.  Phys. D - Applied Physics
{\bf{20}}, 82 (1987).

\bibitem{HN81} S. H\r{a}rd and R. D. Neuman,  J. Colloid Int. Sci. {\bf{83}} , 315 (1981).

\bibitem{WE88}  P. J. Winch and J. C. Earnshaw, J.  Phys. E -
Scientific Instruments {\bf{21}}, 287 (1988).

\bibitem{Goo81}  F. C. Goodrich,
Proc. Royal Society of London - Series A {\bf{374}}, 341 (1981).

\bibitem{DG77} M. Daoud and P. G. d. Gennes, J.  Phys. (France) {\bf{38}}, 85
(1977).

\bibitem{Vil?} R.Vilanove and F.Rondelez, Phys. Rev. Lett.
{\bf{45}}, 1502 (1980)

\bibitem{Ear90}  J. C. Earnshaw, R. C. McGivern, A. C. McLaughlin,
and P. J. Winch, Langmuir {\bf{6}}, 649 (1990).

\bibitem{Buz98}  D. M. A. Buzza,
J. L. Jones, T. C. B. McLeish, and R. W. Richards, J. Chem. Phys.
{\bf{109}}, 5008 (1998).

\bibitem{EM88}   J. C. Earnshaw
and R. C. McGivern,  J. Colloid Int. Sci. {\bf{123}}, 36 (1988).



\bibitem{Imb99}  J.B.Imbert, J.M.Victor, N.Tsunekawa, Y.Hiwatari, Phys. Lett.
A {\bf{258}}, 92 (1999).

\bibitem{Hig91}  P.G.Higgs and J.F.Joanny, J. Chem. Phys. {\bf{94}}, 1543 (1991).

\bibitem{Joa00}  J.F.Joanny, M.Castelnovo and R.Netz,  J. Phys.: Condens. Matter {\bf{12}}, A1 (2000).


\bibitem{LT72}  J. Lucassen and M. v. d. Tempel,  J. Colloid Int. Sci. {\bf{41}}, 491 (1972).

\bibitem{Hen92}  M. Hennenberg, X.-L. Chu, A. Sanfeld, and M. G. Verarde,
 J. Colloid Int. Sci. {\bf{150}}, 7 (1992).


\bibitem{MOR99}  F. Monroy, F. Ortega,
and R. G. Rubio, J. Phys. Chem. B {\bf{103}}, 2061 (1999).

\bibitem{Cla94} D. C. Clark, A. R. Mackie, P. J. Wilde, and D. R. Wilson,
Faraday Discussions {\bf{98}}, 253 (1994).



\bibitem{Bro99}  A. S. Brown, R. W. Richards, D. M. A. Buzza, and T. C.
B. McLeish, Faraday Discussions {\bf{112}}, 309 (1999).

\bibitem{RRT96}  R. W. Richards, B. R. Rochford, and M. R. Taylor,
Macromolecules {\bf{29}}, 1980 (1996).

\bibitem{Miz00}  D.Mizuno, K.Hattori, N.Sakamoto, K.Sakai, K.Takagi, Langmuir {\bf{16}}, 643 (2000).

\bibitem{MOR98}  F. Monroy, F. Ortega, and R. G. Rubio, Phys. Rev. E {\bf{58}},
7629 (1998).

\bibitem{CH01}  P.Cicuta and I.Hopkinson, in preparation.

\bibitem{Nos95}  B. A. Noskov, Colloid  Polym. Sci. {\bf{273}}, 263 (1995).

\end{thebibliography}
\end{document}